\documentclass[english,doublespacing]{elsart}
\usepackage[T1]{fontenc}
\usepackage[latin1]{inputenc}
\usepackage{graphicx}
\usepackage[numbers]{natbib}

\makeatletter

\providecommand{\tabularnewline}{\\}

\usepackage{subfigure}

\usepackage{color}

\makeatletter

\journal{J. Alloys Compd.}
\usepackage[nomarkers]{endfloat}

\usepackage{pslatex}

\makeatother

\usepackage{pslatex}
\makeatother
\begin{document}
\begin{frontmatter}

\title{Structure and optical properties of $\alpha$- and $\gamma$- cerium
sesquisulfide}

\author{Ren\'{e} Windiks{$^{1}$ and Erich Wimmer} }

\address{\emph{Materials Design s.a.r.l.}\\
 44, avenue F.-A. Bartholdi\\
 72000 Le Mans, France}

\thanks{To whom the correspondence should be addressed. Current address:~~\\
 Dorfstrasse 63B, 5417 Untersiggenthal, Switzerland~~\\
 Phone: +41 56 282 04 64, Email: rwindiks@materialsdesign.com}

\author{Leonid Pourovskii}

\address{\emph{Ecole Polytechnique, Centre de Physique Th\'{e}orique}\\
 91128 Palaiseau Cedex, France\emph{}\\
\emph{and}\\
\emph{Materials Design s.a.r.l.}\\
 44, avenue F.-A. Bartholdi\\
 72000 Le Mans, France}

\author{Silke Biermann and Antoine Georges}

\address{\emph{Ecole Polytechnique, Centre de Physique Th\'{e}orique}\\
 91128 Palaiseau Cedex, France}

\begin{abstract}
Structural and electronic properties of the $\alpha$- and $\gamma$-phases
of cerium sesquisulfide, Ce$_{2}$S$_{3}$, are examined by first-principles
calculations using the GGA+U extension of density functional theory.
The strongly correlated $f$-electrons of Ce are described by a Hubbard-type
on-site Coulomb repulsion parameter. A single parameter of $U'=4$
eV yields excellent results for crystal structures, band gaps, and
thermodynamic stability for both Ce$_{2}$S$_{3}$ allotropes. This
approach gives insights in the difference in color of brownish-black
$\alpha$-Ce$_{2}$S$_{3}$ and dark red $\gamma$-Ce$_{2}$S$_{3}$.
The calculations predict that both Ce$_{2}$S$_{3}$ modifications
are insulators with optical gaps of 0.8 eV ($\alpha$-phase) and 1.8
eV ($\gamma$-phase). The optical gaps are determined by direct electronic
excitations at $\mathbf{k}=\Gamma$ from localized and occupied Ce $4f$-orbitals
into empty Ce $5d$-states. The $f$-states are situated between the
valence and conduction bands. The difference of 1 eV between the optical
gaps of the two Ce$_{2}$S$_{3}$ modifications is explained by different
coordinations of the cerium cations by sulfur anions. For both Ce$_{2}$S$_{3}$
modifications the calculations yield an effective local magnetic moment
of 2.6 $\mu_{B}$ per cerium cation, which is in agreement with measurements.
The electronic energy of the $\alpha$-phase is computed to be 6 kJ~mol$^{-1}$
lower than that of the $\gamma$-phase, which is consistent with the
thermodynamic stability of the two allotropes. 
\end{abstract}
\end{frontmatter}

\pagebreak

\section{Introduction}

\begin{flushleft}Rare earth (RE) elements are essential ingredients
in a large variety of materials of vital technological, economic,
and ecological importance. This includes energy-efficient lighting
systems, plasma displays, medical imaging, automotive catalysts, permanent
magnets, and pigments \citep{haxelhedrickorris-2002}. These applications
depend on the unique structural, electronic, optical, magnetic, and
chemical properties of RE compounds. The origin for many of these
properties is the presence of localized $4f$-electrons interacting
with the valence electrons. In this context, the present work focuses
on cerium sesquisulfide (Ce$_{2}$S$_{3}$), which is used as an environmentally
responsible pigment with high temperature stability \citep{berte-2002}.
Beyond this specific application, Ce$_{2}$S$_{3}$ can be seen as
a representative of a large class of RE compounds containing $4f$-electrons.
Therefore, the approach and results of the present work has implications
for many RE-containing compounds and their applications.\par\end{flushleft}

\begin{flushleft}The localization of the $4f$-electrons involves
strong correlation effects which are not captured by current standard
methods within density functional theory (DFT). The result is that these
DFT calculations fail to reproduce, e.g. the non-metallic ground state
of Ce$_{2}$S$_{3}$ \citep{perrinwimmer544-1996,zhukovmauricotgressierevain1282-1997}
and overestimate the mass density of crystal structures of other cerium
compounds \citep{fabrisgironcolibaronivicariobalducci71-2005,kresseblahasilvagandugliapirovano72-2005,fabrisgironcolibaronivicariobalducci72-2005}.
Hence, approaches beyond conventional DFT methods are required to
describe Ce$_{2}$S$_{3}$ realistically from first principles. The
present work demonstrates the usefulness of a DFT method extended
by a Hubbard-type on-site Coulomb repulsion parameter for treating
the Ce $4f$-electrons (the GGA+U approach).\par\end{flushleft}

\begin{flushleft}Cerium sesquisulfide exists in three allotropic forms
\citep{marrotmossettrombemacaudieremaestro259-1997,romeromossettrombemacaudiere78-1997}.
The $\alpha$-phase is the low-temperature modification and has a
brownish-black (rusty) hue \citep{picondomangeflahautguittardpatrie2-1960}.
Above $\sim$900$^{\circ}$C $\alpha$-Ce$_{2}$S$_{3}$ transforms
irreversibly into the dark red (maroon) colored $\beta$-phase. The
$\beta$-modification has been described as an oxysulfide with the
formula Ce$_{10}$S$_{15-x}$O$_{x}$ \citep{schleidlauxmann625-1999}.
The oxygen can be partialy replaced by sulfur. However, a complete
substitution appears to be impossible \citep{schleidlauxmann625-1999},
despite other assertions of Picon \emph{et al.} \citep{picondomangeflahautguittardpatrie2-1960}.
The structure of the $\beta$-phase has been described as complex,
but has not been resolved. For these reasons, the $\beta$-phase is
not considered in the present computational work. The high-temperature
allotrope, $\gamma$-Ce$_{2}$S$_{3}$, is preferentially formed between
1100$^{\circ}$C and 1200$^{\circ}$C. However, $\gamma$-Ce$_{2}$S$_{3}$
can be stabilized at temperatures below 800$^{\circ}$C \citep{trombeverelst323324-2001,berte-2002}.
The melting point of the $\gamma$-phase is at $\sim$2100$^{\circ}$C
\citep{picondomangeflahautguittardpatrie2-1960,hendersonmuramotolohgruber479-1967}.\par\end{flushleft}

\begin{flushleft}The $\gamma$-phase of Ce$_{2}$S$_{3}$ is a nontoxic
inorganic high performance pigment of dark red (burgundy) hue, which
is used on an industrial scale in the coloration of plastics, paints,
and coatings \citep{berte-2002}. Optical studies \citep{dagysbabonaspukinskas5111-1995,witzhugueninlafaitduponttheye794-1996}
suggest that the dark red color of $\gamma$-Ce$_{2}$S$_{3}$ arise
from dipole-allowed excitations of electrons from $4f$-states into
the empty $5d$-states ($4f^{1}5d^{0}\rightarrow4f^{0}5d^{1}$) of
the same cerium cation. The three modifications of Ce$_{2}$S$_{3}$
have an insulating nature, as is already clear from their color.\par\end{flushleft}

\begin{flushleft}The magnetic susceptibility of $\alpha-$Ce$_{2}$S$_{3}$
and $\gamma$-Ce$_{2}$S$_{3}$ at room temperature are similar \citep{picondomangeflahautguittardpatrie2-1960}:
2.9$\times10^{7}$ ($\alpha$-phase) and 2.7 $\times10^{7}$ ($\gamma$-phase),
in SI units. The $\gamma$-phase exhibits ferromagnetism below 3~K
and the measured effective magnetic moment of each cerium cation is
$\mu_{eff}=2.5$ $\mu_{B}$ \citep{beckerfeldhauswesterholtmethfessel6-1977,hotahergrubergschneidnder263-1982,moerkekaldiswachter335-1986}.
The magnetic moment is equivalent to that of an isolated Ce$^{3+}$
ion which has the electron configuration {[}Xe]$4f^{1}$ and the ground
state $^{2}F_{5/2}$ \citep{greenwoodearnshaw-1990}.
Thorough literature searches did not yield anything about a magnetic
ordering of $\alpha$- Ce$_{2}$S$_{3}$ at low temperatures. The
$\alpha$- and $\gamma$- phases of Gd$_{2}$S$_{3}$ have the same
magnetic ordering at low temperatures \citep{ebisuiijimaiwasanagata65-2004}.
Moreover, $\alpha$-Gd$_{2}$S$_{3}$ and $\gamma$- Gd$_{2}$S$_{3}$
are isomorphic to $\alpha$- Ce$_{2}$S$_{3}$ and $\gamma$-Ce$_{2}$S$_{3}$,
respectively. For these reasons, we shall assume in the following
that both $\alpha$- Ce$_{2}$S$_{3}$ and $\gamma$-Ce$_{2}$S$_{3}$
are ferromagnetic at low temperatures.\par\end{flushleft}

\begin{flushleft}The conduction band (cb) of $\gamma$-Ce$_{2}$S$_{3}$
is mainly formed by empty Ce $5d$-states while the valence band (vb)
consists predominantly of sulfur $3p$-orbitals \citep{perrinwimmer544-1996,zhukovmauricotgressierevain1282-1997,dagysbabonaspukinskas5111-1995,witzhugueninlafaitduponttheye794-1996,babonasdagyspukinskas153-1989}.
Due to their rather extended character and large energy dispersion
Ce $6s$-states contribute to both the vb and the cb. The measured
gap between the upper edge of the vb and the lower edge of the cb
of $\gamma$-Ce$_{2}$S$_{3}$, $E_{g}$, is between 2.7 eV \citep{witzhugueninlafaitduponttheye794-1996,golubkovprokofievshelykh376-1995,prokofievshelykhmelekh242-1996}
and 3.3 eV \citep{witzhugueninlafaitduponttheye794-1996}. In agreement
with group-theoretical arguments \citep{zhuzekamarzinkarinsidorinshelykh21-1979}
the smaller band gap has been assigned to phonon-assisted (indirect)
electronic excitations. The larger band gap is associated with direct
electronic transitions. The temperature dependence of $E_{g}$ between
250 and 350 K is approximately $-$0.1 $\mu$eV~K$^{-1}$, which has
been derived from $\gamma$-La$_{2}$S$_{3}$ and $\gamma$-Dy$_{2}$S$_{3}$
\citep{zhuzekamarzinkarinsidorinshelykh21-1979}. Both compounds have
the same structural topology as $\gamma$-Ce$_{2}$S$_{3}$. The corresponding
values of typical semiconductors like CdS and CdSe are around $-$4 $\mu$eV~K$^{-1}$.\par\end{flushleft}

\begin{flushleft}Photoconductivity spectra \citep{mauricotbullotweryevian313-1996}
as well as measurements of the photoluminescence \citep{mauricotdexpertghysevian69-1996}
of $\gamma$-Ce$_{2}$S$_{3}$ indicate that the occupied $4f$-states
of cerium cations are energetically located between the vb and the
cb. The gap between the $4f$-states and the cb obtained from these
measurements is in agreement with the optical gap of 1.8--2.0 eV derived
from the dielectric function \citep{dagysbabonaspukinskas5111-1995,witzhugueninlafaitduponttheye794-1996}.
Up to now, photoemission spectroscopy studies of $\gamma$-Ce$_{2}$S$_{3}$
have not been able to resolve precisely the energetic position of
the occupied Ce $4f$-states with respect to the vb \citep{kaciulislatisenkaplesanovas2512-1991,witzhugueninlafaitduponttheye794-1996}.
However, the measurements of Mauricot \emph{et al.} \citep{mauricotbullotweryevian313-1996}
reveal that the occupied Ce $4f$-states are energetically 0.8 eV
above the vb.\par\end{flushleft}

\begin{flushleft}Earlier band structure calculations within DFT in
the local density approximation (DFT-LDA) on model structures for
$\gamma$-Ce$_{2}$S$_{3}$ demonstrated qualitatively the importance
of $4f$-electrons for the optical properties \citep{perrinwimmer544-1996,zhukovmauricotgressierevain1282-1997}.
However, these calculations have several limitations, which hamper
quantitative predictions. These limitations mostly stem from the well
known inability of DFT-LDA methods to describe correctly the localized
$4f$-states. First, when the $4f$-states are treated in the valence,
DFT-LDA calculations predict a metallic (Kohn-Sham) band structure
for Ce$_{2}$S$_{3}$, with $f$-states at the Fermi level, in contradiction
with the insulating nature of this material. The fact that, within
DFT-LDA, these $f$-bands are found to be split off from the sulfur
$p$-bands is indeed suggestive that the optical transitions from
$4f$ to $5d$ bands are playing an important role in the optical
properties. But it is clearly not possible to go beyond this qualitative
observation in a pure DFT-LDA framework, let alone to reach a quantitative
description of the optical gaps of Ce$_{2}$S$_{3}$ derivatives \citep{perrinwimmer544-1996,zhukovmauricotgressierevain1282-1997}.
Second, the models used to describe $\gamma$-Ce$_{2}$S$_{3}$ in
Refs.~\citep{perrinwimmer544-1996,zhukovmauricotgressierevain1282-1997}
did not capture the electrostatics of the real compound adequately.
The charges of the cations in the model significantly deviate from
that of trivalent cerium. The deviation was one of the factors for
the DFT-LDA calculations to underestimate the energy difference between
the $4f$-states and the vb. The result was an overestimation of the
overlap between the $4f$-states and the vb and an artificially large
splitting of band states on top of the vb \citep{perrinwimmer544-1996}.
Third, the previous LDA-DFT calculations did not perform structural
relaxation studies or attempt to predict crystal structures. However,
the electronic structure and, hence, optical properties are very sensitive
to interatomic distances. Furthermore, it is widely documented that
DFT-LDA and even the generalized gradient approximation (GGA), an
extensions of DFT-LDA, severely overestimate the mass density of $f$-electron
compounds when the $f$-states are treated in the valence \citep{fabrisgironcolibaronivicariobalducci71-2005,savrasovkotliar8416-2000,skorodumovaahujasimakabrikosovjohanssonlundqvist64-2001}.
Our own experience with alkali-doped cerium sesquisulfides is that
GGA calculations underestimate the unit cell volume of these compounds
by as much as $6\%$ \citep{windikspourovskiygeorgeswimmer-2006}.\par\end{flushleft}

\begin{flushleft}The main goal of the present study is to gain a deeper
understanding of the relations between the chemical composition, the
crystal structure, and the optical properties of both $\alpha$-Ce$_{2}$S$_{3}$
and $\gamma$-Ce$_{2}$S$_{3}$. Our study addresses both the crystal
structure as well as the electronic properties quantitatively within
a single computational framework, namely the GGA+U approach. Analogue
to the LDA+U method \citep{anisimovaryasetiawanlichtenstein9-1997},
the GGA+U approach uses the DFT framework but extends the GGA description
of exchange-correlation effects in the localized $4f$-shell by a
many-body interaction term. This term is then treated in a mean-field
like manner, allowing for a simple description of strong correlation
effects within the Ce $4f$-states. The GGA+U approach allows for a
full optimization of the crystallographic parameters of Ce$_{2}$S$_{3}$,
and captures the role of the $4f$-electrons in the optical transitions.
As we shall see, Ce$_{2}$S$_{3}$ is found to be an $f$-electron Mott
insulator, in which the Ce $4f$-states are split into lower occupied
and upper unoccupied sub-bands by the strong on-site Coulomb repulsion.
Optical transitions responsible for the color of these compounds are
between the lower unoccupied Ce $4f$-states and the bottom of the
Ce $5d$-like cb.\par\end{flushleft}

\section{Crystal structures \label{sec:CrystStrcts}}

\begin{flushleft}%
\begin{figure*}
\includegraphics[width=1.0\columnwidth,keepaspectratio]{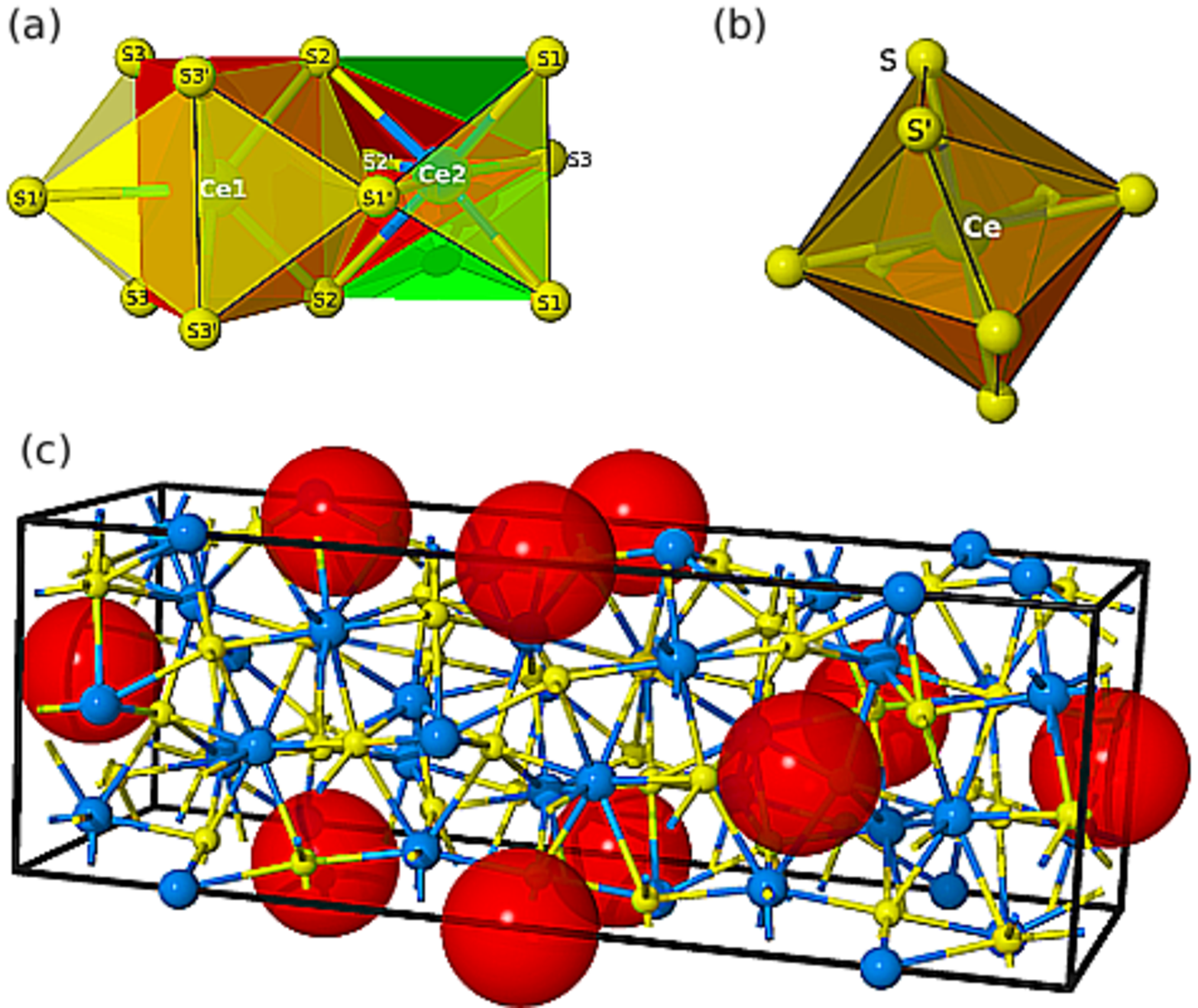}

\caption{\label{fig:Ce2S3.StrctFeatures}(color online) Structural features
of cerium sesquisulfide allotropes. Blue and yellow spheres represent
cerium cations and sulfur anions, respectively. (a) Coordination polyhedra
of the symmetry inequivalent cerium cations Ce1 and Ce2 of $\alpha$-Ce$_{2}$S$_{3}$.
(b) Coordination polyhedron of Ce cations in $\gamma$-Ce$_{2}$S$_{3}$.
The sulfur anions S and S' are within and out of the drawing plane,
respectively. (c) Smallest conventional unit cell of $\gamma$-Ce$_{2}$S$_{3}$with
a stoichiometric amount of cerium vacancies (red spheres). The composition
of the super cell is Ce$_{32}$S$_{48}$ and has the highest symmetry
possible (space group $I\bar{4}2d$).}
\end{figure*}
\par\end{flushleft}

\begin{flushleft}The $\alpha$-phase of Ce$_{2}$S$_{3}$ crystallizes
in an orthorhombic lattice (crystallographic space group $Pnma$).
The unit cell contains four formula units with two and three symmetry
inequivalent cerium and sulfur ions, respectively \citep{schleidlauxmann625-1999}.
Figure \ref{fig:Ce2S3.StrctFeatures}(a) illustrates two different
kinds of interlocked Ce--S skeletons which are the coordination polyhedra
of the four Ce1 and the four Ce2 cations. The coordination polyhedra
(Ce1)S$_{8}$ and (Ce2)S$_{7}$ have the shapes of a bi-capped trigonal
prism and a rather distorted mono-capped octahedron, respectively.
The corners of the trigonal prism are S2, S3, and S3' whereas the
vertices of the two caps are the sulfur ions S1'. The cap tip of the
octahedron is the sulfur ion S2'. Two of the three symmetry inequivalent
sulfur anions (S2 and S3) are bound to five cerium cations. The cations
form a rectangular pyramid. The five cerium cations encompassing the
remaining symmetry inequivalent sulfur anion (S1) form an axial distorted
trigonal bipyramid. The Ce--S distances of the two coordination polyhedra
are in the range of 2.85--3.14 Å with average distances of 3.01 Å
{[}(Ce1)S$_{8}$] and 2.93 Å {[}(Ce2)S$_{7}$].\par\end{flushleft}

\begin{flushleft}The crystal structure of $\gamma$-Ce$_{2}$S$_{3}$
has a body-centered cubic lattice. It is isomorphic to the Th$_{3}$P$_{4}$
structure and contains vacancies on the cationic sublattice \citep{schleidlauxmann625-1999,zachariasen2-1949,carter5-1972}.
The cerium cations are located on the 12$a$ positions which have
the point group symmetry $S_{4}$; the sulfur anions occupy sites
with $C_{3}$ symmetry (16$c$ positions). Each cerium cation is coordinated
by eight sulfur anions forming a triangular dodecahedron as depicted
in Fig.~\ref{fig:Ce2S3.StrctFeatures}(b), with two different nearest-neighbor
distances \citep{mauricotgressierevianbrec223-1995}: $r$(Ce--S)=2.89 Å
and $r$(Ce--S')=3.09 Å. Each sulfur anion is located on a threefold
axis and surrounded by six nearest cerium cations, provided all 12$a$
positions are occupied. The six cations form a distorted octahedron.
However, when all the 16$c$ and 12$a$ position are occupied with
cations and anions, respectively, the stoichiometry would be Ce$_{3}$S$_{4}$.
In order to achieve the stoichiometry of $\gamma$-Ce$_{2}$S$_{3}$
one out of nine cations has to be removed and vacancies on the cerium
sublattice, V$_{\mathrm{Ce}}$, have to be created \citep{carter5-1972}:\[
3\times[\mathrm{Ce}_{3}\mathrm{S}_{4}]\,\equiv\,\mathrm{Ce}_{9}\mathrm{S}_{12}\stackrel{-\mathrm{Ce}}{\longrightarrow}\mathrm{Ce_{8}V_{Ce}S_{12}}\,\equiv\,\mathrm{4\times[Ce_{2}S_{3}}]\]
 Note that both, $\gamma$-Ce$_{2}$S$_{3}$ and Ce$_{3}$S$_{4}$
have very similar lattice parameters differing only by 0.06\% \citep{zachariasen2-1949,mauricotgressierevianbrec223-1995}.
The tiny volume difference is evidence for a rather rigid Ce-S framework.\par\end{flushleft}

\begin{flushleft}Raman spectra of several RE sesquisulfides in the
$\gamma$-phase give evidence for randomly distributed vacancies \citep{knightwhite463-1990}.
However, simple electrostatic energy calculations predicted that a
completely random distribution of the voids on the cerium sublattice
yields an energetically less stable ground state than a vacancy ordering
such that the resulting crystal structure of $\gamma$-Ce$_{2}$S$_{3}$
has the symmetry of the space group $I\bar{4}2d$ \citep{carter5-1972}.
A highly symmetric ordering of the vacancies is likely in well-annealed
$\gamma$-phases whereas rapidly quenched samples might have a random
distribution of vacancies \citep{carter5-1972}. Vacancies in the
sulfur sublattice are rather unlikely \citep{kolesovkamarzinsokolov384-1997}.\par\end{flushleft}

\begin{flushleft}Figure \ref{fig:Ce2S3.StrctFeatures}(c) depicts
the smallest conventional crystallographic supercell that accommodates
the stoichiometry of $\gamma$-Ce$_{2}$S$_{3}$. The supercell is
three times as big as the conventional unit cell of Ce$_{3}$S$_{4}$
and has the composition $4\times${[}Ce$_{8}$V$_{\mathrm{Ce}}$S$_{12}$].
The distances of a cerium cation to its first, second and third nearest
cations in $\gamma$-Ce$_{2}$S$_{3}$ are 7.5, 4.8 and 4.0 Å \citep{mauricotgressierevianbrec223-1995}.
In this structure the voids on the cation sublattice create three
symmetry inequivalent cerium cations. The supercell can be transformed
into a primitive unit cell half as large as the conventional cell
shown in Fig. \ref{fig:Ce2S3.StrctFeatures}(c).\par\end{flushleft}

\section{Computational Method}

\begin{flushleft}All GGA+U calculations have been performed with the
\emph{Vienna ab initio simulation package} (VASP) \citep{kressefurthmueller54-1996,kressefurthmueller6-1996,kressejoubert593-1999,rohrbachhafnerkresse69-2004,paierhirschlmarsmankresse122-2005}
as integrated in the MedeA environment \citep{medea2}. VASP uses
the projector-augmented wave (PAW) method \citep{bloechl5024-1994}
in the implementation of Kresse and Joubert \citep{kressejoubert593-1999}.
Exchange- and correlation effects are described within the GGA by
the density functional of Perdew, Burke and Ernzerhof \citep{perdewburkeernzerhof7718-1996}.
The PAW potentials take into account 12 valence electrons of each
Ce ($5s^{2}5p^{6}6s^{2}5d^{1}4f^{1}$) and six valence electrons of
each S ($3s^{2}3p^{4}$) atom. The plane wave energy cutoff has been
set to 333 eV and 299 eV for $\alpha$-Ce$_{2}$S$_{3}$ and $\gamma$-Ce$_{2}$S$_{3}$,
respectively.\par\end{flushleft}

\begin{flushleft}Within the present GGA+U approach \citep{rohrbachhafnerkresse69-2004,dudarevbottonsavrasovhumphreyssutton573-1998}
the strong electron correlation among the $4f$-states is described
with the parameter $U'=U-J$. In the latter term, $U$ is the spherically
averaged screened Coulomb repulsion energy required for adding an
extra electron into the Ce $4f$-states and $J$ represents the corresponding
screened exchange energy. The GGA+U one-electron potential is defined
as $V_{ij}^{\sigma}=\partial E_{GGA}/\partial\rho_{ij}^{\sigma}+U'(0.5\delta_{ij}-\rho_{ij}^{\sigma})$,
whereby $E_{GGA}$ is the GGA electronic enenergy, $\rho_{ij}^{\sigma}$
is the density matrix of $f$-electrons with spin $\sigma$ , and
$\delta_{ij}$ is the delta function. The double counting correction
applied is equivalent to that of the fully-localized limit. The case
$U'=0$ is the limit of pure GGA.\par\end{flushleft}

\begin{flushleft}The parameter $U'$ can be calculated from first
principles without knowledge of experimental data \citep{cococcionigironcoli71-2005}.
In this work however, instead of performing such calculations (which
are computationally heavy because of the large numbers of atoms in
the unit cell), we have investigated whether $U'$ can be chosen in
such a way to reproduce with reasonable accuracy \textit{both} the
experimentally measured unit cell volume, as well as the optical gap
of the $\gamma$-Ce$_{2}$S$_{3}$ allotrope. We found that the value
$U'=4$ eV achieves this goal. This value is in agreement with Coulomb
repulsion parameters calculated for other cerium compounds \citep{herbstwatsonwilkens178-1978,fabrisvicariobalduccigironcolibaroni109-2005}.\par\end{flushleft}

\begin{flushleft}The Kohn-Sham equations are solved self-consistently
in their spin-polarized form employing an iterative matrix diagonalization
(blocked Davidson scheme) and the spin interpolation of Vosko et al.
\citep{voskowilknusair58-1980}. The convergence criterion to obtain
the energy is $10^{-5}$~eV. The integration over the first Brillouin-zone
is performed using symmetry adapted Monkhorst-Pack grids \citep{monkhorstpack1312-1976}
of $\mathbf{k}$-point sets of various size. To calculate energies,
energy gradients (atomic forces as well as the stress tensors), and
charge densities the $\mathbf{k}$-point sets are $3\times7\times3$
($\alpha$-Ce$_{2}$S$_{3}$) and $3\times3\times3$ ($\gamma$-Ce$_{2}$S$_{3}$).
All $\mathbf{k}$-point sets as well as the plane wave energy cutoff
has been determined within a convergence test to yield to a numerical
precision of $5\times10^{-3}$ eV.\par\end{flushleft}

\begin{flushleft}The crystal structures of $\alpha$-Ce$_{2}$S$_{3}$
and $\gamma$-Ce$_{2}$S$_{3}$ as described in Sec. \ref{sec:CrystStrcts}
have been optimized such that atomic positions and the crystallographic
cell parameters are relaxed simultaneously. The optimization procedure
has two steps. In the first step, the conjugate-gradient algorithm
is used and the optimization stops when all energy gradients are smaller
than $0.02$~eV~Å$^{-1}$. Then the crystal structures are further
refined within a quasi-Newton algorithm with a very tight convergence
criterion of $0.002$~eV~Å$^{-1}$.\par\end{flushleft}

\begin{flushleft}Energy gradients have been calculated with a Gaussian
integration method and a smearing of $\sigma=10^{-3}$~eV. The linear
tetrahedron method with Blöchl corrections has been used to obtain
accurate energies and the self-consistent charge densities. (This
integration scheme is inappropriate for calculating energy gradients.)
The self-consistent charge densities are used to calculate the band
structures and the densities of states (DOS's). All DOS's have been
calculated using the linear tetrahedron method without corrections
and the same $\mathbf{k}$-point sets applied to calculate the energies
(see above).\par\end{flushleft}

\begin{flushleft}Since the calculations correspond to zero temperature,
the two Ce$_{2}$S$_{3}$ phases adopt the ferromagnetic state. In
all calculations the ferromagnetic state is described with all $f$-electron
spins alligned parallel (spin-up). The band structures and DOS's have
been created from the charge density of the electrons with spin-up.
This is sufficient since in the present approach only the spin-up
states contain occupied $f$ states while the S $3p$-vb and Ce $5d/6s$-conduction
states are similar for both spin orientations.\par\end{flushleft}

\section{Results and discussion}

\subsection{\label{sec:aCe2S3}$\alpha$-Phase}

\begin{flushleft}Table \ref{tab:aCe2S3fopt-strctparam} summarizes
calculated lattice parameters and interionic distances of the two
interlocked (Ce1)S$_{8}$ and (Ce2)S$_{7}$ skeletons shown in Fig.
\ref{fig:Ce2S3.StrctFeatures}(a). As announced above, the chosen
value $U'=4$ eV yields a relative error on the unit cell volume which
is smaller than 1\% in comparison to experiments. A comparison with
the corresponding experimental distances reveals remarkable agreement.
Most of the calculated distances are less than 0.9\% larger than those
determined from x-ray diffraction measurements. Exceptions are the
calculated cell length $c$ and the bond lengths $r$(Ce1--S1') and
$r$(Ce2--S3) which are nearly equal or smaller, respectively, than
the corresponding measured distances. The discrepancies are almost
zero for $c$, $-$0.1\% ($r$(Ce2--S3), and around $-$1.2\% for the longest
of the two Ce1--S1' distances. The calculations tend to strengthen
one of the Ce1--S1' bonds at the expense of the other thereby slightly
decreasing the sulfur coordination of Ce1. Note that the measurements
and the calculations yield for the two distances $r$(Ce1--S1') average
values of 3.12 and 3.11 Å, respectively. The average of the calculated
distances between cerium and sulfur within the (Ce1)S$_{8}$ and (Ce2)S$_{7}$
skeletons are 3.02 and 2.94 Å, respectively. Both distances are only
between 0.4 and 0.7\% larger than the corresponding distances of the
x-ray structure. Apart from $r$(Ce1--Ce1) all other calculated Ce--Ce
distances are larger than the measured values with deviations not
larger than 0.5\%.\par\end{flushleft}

\begin{table*}

\caption{\label{tab:aCe2S3fopt-strctparam} Calculated lattice parameters
and selected interionic distances, Ce1--X and Ce2--X, of $\alpha$-Ce$_{2}$S$_{3}$
compared with data obtained from the x-ray diffraction measurements
of Ref. \citep{schleidlauxmann625-1999}. The notation is according
to that of Fig. \ref{fig:Ce2S3.StrctFeatures}(a). All distances are
given in Å.}

\begin{centering}\begin{tabular}{lrrllrrrrr}
\hline 
\multicolumn{3}{l}{Lattice (space group $Pnma$)}&
&
\multicolumn{6}{c}{Interionic Distances}\tabularnewline
\hline 
&
&
&
&
X&
\multicolumn{2}{c}{Ce1}&
&
\multicolumn{2}{c}{Ce2}\tabularnewline
\hline 
Parameters&
Exp.&
Calc.&
&
&
Exp.&
Calc.&
&
Exp.&
Calc.\tabularnewline
\hline 
$a$&
7.532&
7.565&
&
S1&
&
&
&
2.850&
2.872\tabularnewline
$b$&
4.097&
4.119&
&
S1'&
3.100, &
3.062, &
&
3.040&
3.056\tabularnewline
&
&
&
&
&
3.135&
3.161&
&
&
\tabularnewline
$c$&
15.728&
15.722&
&
S2&
2.983&
2.998&
&
2.883&
2.893\tabularnewline
$\Delta V/V_{exp}$ {[}\%]&
&
0.95&
&
S2'&
4.727&
4.761&
&
2.976&
2.988\tabularnewline
&
&
&
&
S3&
3.006&
3.018&
&
2.897&
2.893\tabularnewline
&
&
&
&
S3'&
2.850&
2.853&
&
4.696&
4.719\tabularnewline
&
&
&
&
Ce1&
4.031&
4.029&
&
4.015&
4.035\tabularnewline
&
&
&
&
Ce2&
4.015&
4.035&
&
4.097&
4.119\tabularnewline
\hline
\end{tabular}\par\end{centering}
\end{table*}

\begin{flushleft}The energy of formation, $E_{form}$, is a measure
for the stability and is the difference between the energy of the
entire system and the energies of the electronic ground state of the
constitutive elements in their standard state. For $\alpha$-Ce$_{2}$S$_{3}$
the calculations yield $E_{form}=-820$ kJ~mol$^{-1}$ Ce$_{2}$S$_{3}$.\par\end{flushleft}

\begin{figure}
\includegraphics[width=0.9\columnwidth]{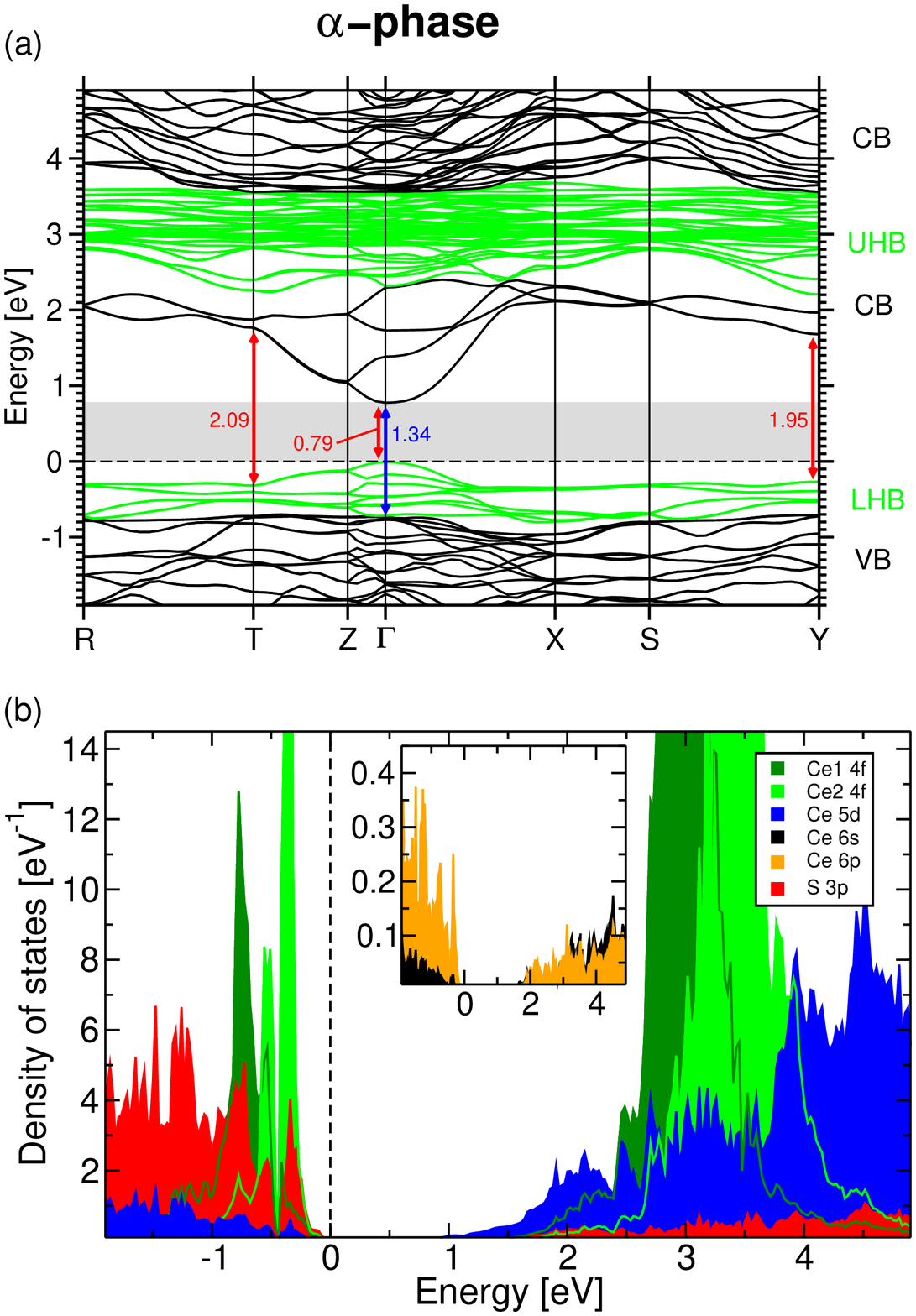}

\caption{\label{fig:aCe2S3fopt.Bands-PDOS} (color online) Electronic structure
of $\alpha$-Ce$_{2}$S$_{3}$ around the Fermi energy; (a) band structure
and (b) partial DOS calculated for the fully optimized crystal structure.
Arrows indicate relevant direct band gaps. The energy axes are labeled
relative to the energy of the highest occupied band state. The latter
is set to zero and is denoted by dashed lines. VB -- valence band,
CB -- conduction band, UHB and LHB -- upper and lower Hubbard bands,
respectively.}
\end{figure}

\begin{flushleft}Figure \ref{fig:aCe2S3fopt.Bands-PDOS} shows the
band structure and the corresponding partial DOS (PDOS) calculated
for the fully optimized crystal structure of $\alpha$-Ce$_{2}$S$_{3}$.
The electronic bands are drawn along the standard pathway within the
irreducible wedge of the Brillioun zone of simple orthorhombic lattices.
The PDOS illustrates that the vb, i.e. the occupied and evenly dispersed
electron bands below approximately $-$0.8 eV, are mainly formed by S $3p$-orbitals
with contributions of Ce $3d$-orbitals. This indicates a significant
covalent contribution to the bonding in $\alpha$-Ce$_{2}$S$_{3}$.\par\end{flushleft}

\begin{flushleft}On top of the vb, i.e. the occupied electron bands
between $-$0.8 eV and zero energy are two kinds of occupied $4f$-orbitals
namely those of the cations Ce1 and Ce2, respectively. In the following
occupied $4f$-states are termed as lower Hubbard bands (LHB's). Both
LHB's have rather large admixtures of S $3p$-states. This indicates
a participation of Ce $4f$-orbitals in the covalent bonding. The PDOS
of Fig. \ref{fig:aCe2S3fopt.Bands-PDOS}(b) shows that the LHB of
the Ce1 cations is located at a lower energy than the LHB of the Ce2
cations. The main reason for the different positions must be the different
sulfur environments of the Ce1 and Ce2 cations as depicted in Fig.
\ref{fig:Ce2S3.StrctFeatures}(a). The calculations yield almost one
$f$-electron localized on each cerium cation. The computed occupation
numbers of the $4f$-states, $n_{f}$, of the two symmetry inequivalent
cerium cations are 1.12 (Ce1) and 1.14 (Ce2). Occupation numbers of
$n_{f}>1$ are consistent with admixtures of S $3p$-states to Ce $4f$-orbitals.
Assuming that the total orbital moment of the electronic ground state
of all cerium cations is $J=5/2$ the effective local magnetic moment
is around 2.6 $\mu_{B}$ per cerium cation. This is consistent with
experiment.\par\end{flushleft}

\begin{flushleft}The LHB's have different features within the chosen
part of the Brillouin zone. On the one hand the LHB's have significant
dispersion (on the scale of $\sim0.5$ eV) between $\mathbf{k}=T$
and the center of the line between $\mathbf{k}=\Gamma$ and $\mathbf{k}=X$
($\Sigma$ line). On the other hand the LHB's are rather flat between
the center of the $\Sigma$ line and $\mathbf{k}=Y$ and along the
line between $\mathbf{k}=R$ and $\mathbf{k}=T$. The LHB's are almost
degenerate at $\mathbf{k}=R$ and $\mathbf{k}=S$.\par\end{flushleft}

\begin{flushleft}The smallest gap between the upper edge of the LHB
of Ce2 and the lower edge of the unoccupied states (cb) is 0.8 eV
and is at $\mathbf{k}=\Gamma$. At the same $\mathbf{k}$-point is
the smallest separation between the vb and the cb (1.34 eV). The cb
has predominantly Ce $5d$-character with small contributions of S $3p$-orbitals.
Like the LHB's but more distinct, the cb exhibits a strong dispersion
between $k=T$ and $k=X$. The PDOS of the cb at its lower edge is
tiny and increases in small steps between 0.7 and 1.8 eV.\par\end{flushleft}

\begin{flushleft}The inset of Fig. \ref{fig:aCe2S3fopt.Bands-PDOS}(b)
depicts the presence of Ce $6s$- and Ce $6p$-orbitals in the relevant
electron bands of Fig. \ref{fig:aCe2S3fopt.Bands-PDOS}(a). Their
contributions within the relevant energy range around the Fermi energy
are relatively small because the $6s$- and $6p$-states are spatially
extended and energetically dispersed. Inside the cb between 2.0 and
4.2 eV are unoccupied Ce $4f$-orbitals which form the upper Hubbard
band (UHB).\par\end{flushleft}

\begin{figure}
\includegraphics[width=0.95\columnwidth]{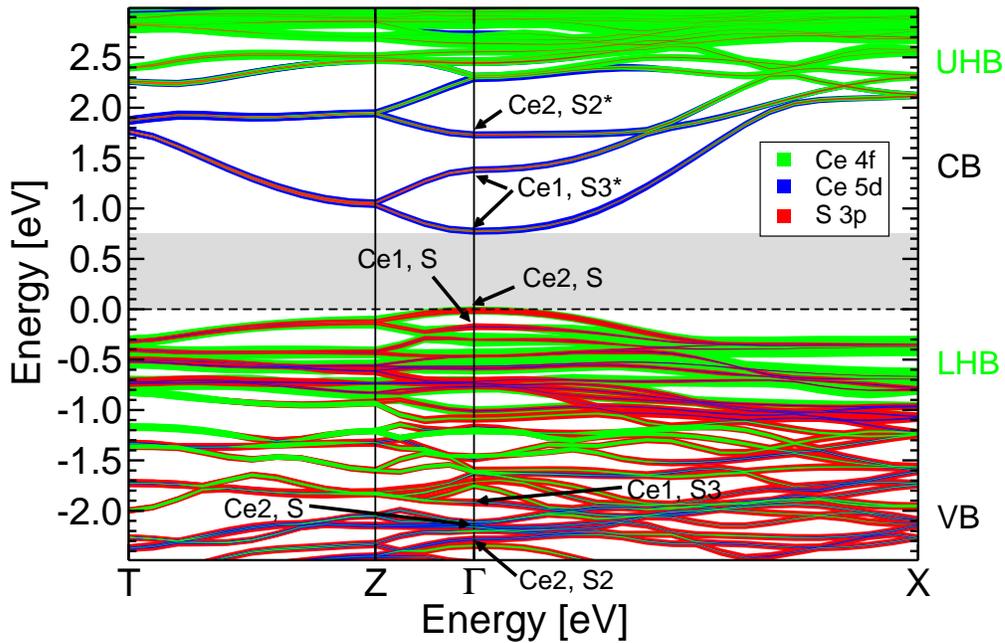}

\caption{\label{fig:aCe2S3fopt.OPBands} (color online) Contributions of atomic
states per $\mathbf{k}$-point to particular electronic bands of $\alpha$-Ce$_{2}$S$_{3}$
between $k=T$ and $X$ as shown in Fig. \ref{fig:aCe2S3fopt.Bands-PDOS}.
The thicker the electronic bands the larger the contribution of particular
atomic states. Significant contributions of specific atomic states
are indicated at $\mathbf{k}=\Gamma$. Asterisks denote anti-bonding
interactions. More details are given in the legend of Fig. \ref{fig:aCe2S3fopt.Bands-PDOS}.}
\end{figure}

\begin{flushleft}The band dispersions of the LHB and the cb are examined
in terms of an orbital-projected band-structure (sometimes called
{}``fat-band'' representation). Figure \ref{fig:aCe2S3fopt.OPBands}
illustrates contributions of atomic states to relevant energy bands
of Fig. \ref{fig:aCe2S3fopt.Bands-PDOS}(a) between $\mathbf{k}=T$
and $\mathbf{k}=X$ at different $\mathbf{k}$-points. The three lowest
unoccupied states at $\mathbf{k}=\Gamma$ are due to anti-bonding
interactions between cerium $5d$- and sulfur $3p$-states. The contribution
of S $3p$-orbitals to the three states is between 5 and 15\% whereby
only states of the atoms S2 and S3 are involved. Admixtures of $3p$-orbitals
of S1 are negligibly small. The corresponding bonding states are located
between $-1.9$ and $-2.3$ eV. The splitting between bonding and
antibonding states of Ce1 and S3 is between 0.7 and 1.4 eV smaller
than that of Ce2 and S2. The smaller splitting and, hence, the distinct
dispersion of the two lowest unoccupied bands is due to smaller overlap
between $5d$- and $3p$-orbitals of Ce1 and S3, respectively.\par\end{flushleft}

\begin{flushleft}The dispersions at the upper edge of the LHB are
also due to admixtures of S $3p$-orbitals to the Ce $4f$-states. The
two highest occupied electronic states of the LHB contain at $\mathbf{k}=\Gamma$
both 44\% of S $3p$-orbitals whereby all sulfur ions contribute equally.
The remaining contributions come from the $4f$-states of Ce1 and
Ce2, respectively (\emph{cf.} Fig. \ref{fig:aCe2S3fopt.OPBands}).
The contributions of all cerium cations is consistent with the small
tails of the $4f$-states in the PDOS of Fig. \ref{fig:aCe2S3fopt.Bands-PDOS}(b)
at zero energy. In fact the entire vb contains contributions of $4f$-states
which decrease towards lower energies.\par\end{flushleft}

\begin{flushleft}The calculated GGA+U band structure yields insight
into the optical properties of $\alpha$-Ce$_{2}$S$_{3}$ and its
color. The dispersion at the lower edge of the cb at $\mathbf{k}=\Gamma$
causes the rather small direct band gap of 0.8 eV. However, this gap
is between states of Ce1 and Ce2. Since the color of the cerium sesquisulfides
is determined by the excitations $4f^{1}5d^{0}\rightarrow4f^{0}5d^{1}$
taking place on the same site we have to distinguish between two optical
transitions. The orbital-projected band-structure of Fig. \ref{fig:aCe2S3fopt.OPBands}
reveals that the smallest $4f$--$5d$ gaps of Ce1 and Ce2 are about
1.0 and 1.7 eV, respectively. Consequently, the reason why $\alpha$-Ce$_{2}$S$_{3}$
has a brownish-black color is twofold. The cerium cations Ce2 absorb
photons with energies larger than 1.7 eV which would lead to a dark
red color. The smaller gap of $\sim$1.0 eV (Ce1 cations) is responsible
that all radiation of the visible range is absorbed. However, the
absorption of radiation with an energy between 0.7 and 1.8 eV (near
infrared and red light) is relatively low because the density of states
is rather small in this region {[}compare Fig. \ref{fig:aCe2S3fopt.Bands-PDOS}(b)].
This effect leads to the black shade. Dark red mixed with black yields
a brown-black hue.\par\end{flushleft}

\subsection{$\gamma$-Phase}

\begin{table*}

\caption{\label{tab:Ce32S48fopt-StrctParam} Comparison of calculated and
measured lattice parameters and selected interionic distances of $\gamma$-Ce$_{2}$S$_{3}$.
The calculated data have been obtained from the fully optimized supercell
of Ce$_{16}$S$_{24}$ as depicted in Fig. \ref{fig:Ce2S3.StrctFeatures}(c).
The experimental data has been taken from Ref. \citep{mauricotgressierevianbrec223-1995}.
All values are given in Å.}

\begin{centering}\begin{tabular}{ccccccc}
\hline 
\multicolumn{3}{c}{Lattice (Quasi cubic)}&
&
\multicolumn{3}{c}{Distances}\tabularnewline
\hline 
Parameter&
Exp.&
Calc.&
&
&
Exp.&
Calc.\tabularnewline
\hline 
$a$&
8.637&
8.677&
&
$r_{min}$(Ce$\cdots$Ce)&
4.040&
3.958\tabularnewline
$c$&
8.637&
8.641~$^{a)}$&
&
$\langle r$(Ce--S)$\rangle$~$^{c)}$&
2.896&
2.898\tabularnewline
$\Delta V/V_{exp}\times100$&
&
0.97&
&
$\langle r$(Ce--S')$\rangle$~$^{c)}$&
3.087&
3.086\tabularnewline
&
&
&
&
$r$(Q--S)~$^{b)}$&
2.896&
2.988\tabularnewline
&
&
&
&
$r$(Q--S')~$^{b)}$&
3.087&
3.208\tabularnewline
&
&
&
&
$r_{min}$(Q$\cdots$Q)~$^{b)}$&
7.785&
7.779\tabularnewline
\hline
\end{tabular}\par\end{centering}

$^{a)}$~{\scriptsize $c/3$ of the tetragonal supercell of Fig.
\ref{fig:Ce2S3.StrctFeatures}(c).}{\scriptsize \par}

$^{b)}$~{\scriptsize Q is the center of the voids on the cation
sublattice.}{\scriptsize \par}

$^{c)}$~{\scriptsize The notation is according to Fig. \ref{fig:Ce2S3.StrctFeatures}(b).} 
\end{table*}

\begin{flushleft}Table \ref{tab:Ce32S48fopt-StrctParam} summarizes
lattice parameters and selected interionic distances calculated for
$\gamma$-Ce$_{2}$S$_{3}$ and compares the values with that of the
x-ray structure of Ref.~\citep{mauricotgressierevianbrec223-1995}.
Again, the chosen value of $U'$ yields a calculated unit-cell volume
which is less than $1\%$ larger than that determined from experiment.
The deviation between the calculated and measured average Ce--S bond
lengths is negligibly small. The smallest calculated Ce--Ce distance
which is 2\% smaller than that determined from the x-ray structure.
The reason for the cell expansion is a growth of the voids located
on the cation sublattice {[}\emph{cf.} red spheres of Fig. \ref{fig:Ce2S3.StrctFeatures}(c)].
The calculated structure has distances between the center of the voids,
Q, and the nearest sulfur anions which are between 3.1\% and 4.0\%
larger than those of the experimental structure.\par\end{flushleft}

\begin{flushleft}The $I\bar{4}2d$ structure of Fig. \ref{fig:Ce2S3.StrctFeatures}(c)
has three symmetry inequivalent cerium cations and sulfur anions,
respectively. Two of the symmetry inequivalent cerium cations have
clearly distinguishable bond distances of $r$(Ce--S)$\sim$2.90 and $r$(Ce--S')$\sim$3.10
Å. However, the third kind of cations has an almost continuous spectrum
of bond length to its eight nearest sulfur anions in the range of
2.90 and 3.30 Å. Due to the vacancies two of the three symmetry inequivalent
sulfur anions are bound to five cerium cations only. All the penta-coordinated
sulfur anions have three Ce--S bonds and two Ce--S' bonds {[}\emph{cf.}
Fig. \ref{fig:Ce2S3.StrctFeatures}(b)]. The six-fold coordinated
sulfur anions have Ce--S distances, which are between 1 and 2\% larger
than the calculated average bond lengths of Tab. \ref{tab:Ce32S48fopt-StrctParam}.
The calculated energy of formation for the $\gamma$-phase is $\Delta E_{form}=-814$
kJ~mol$^{-1}$ Ce$_{2}$S$_{3}$.\par\end{flushleft}

\begin{figure}

\includegraphics[width=0.9\columnwidth]{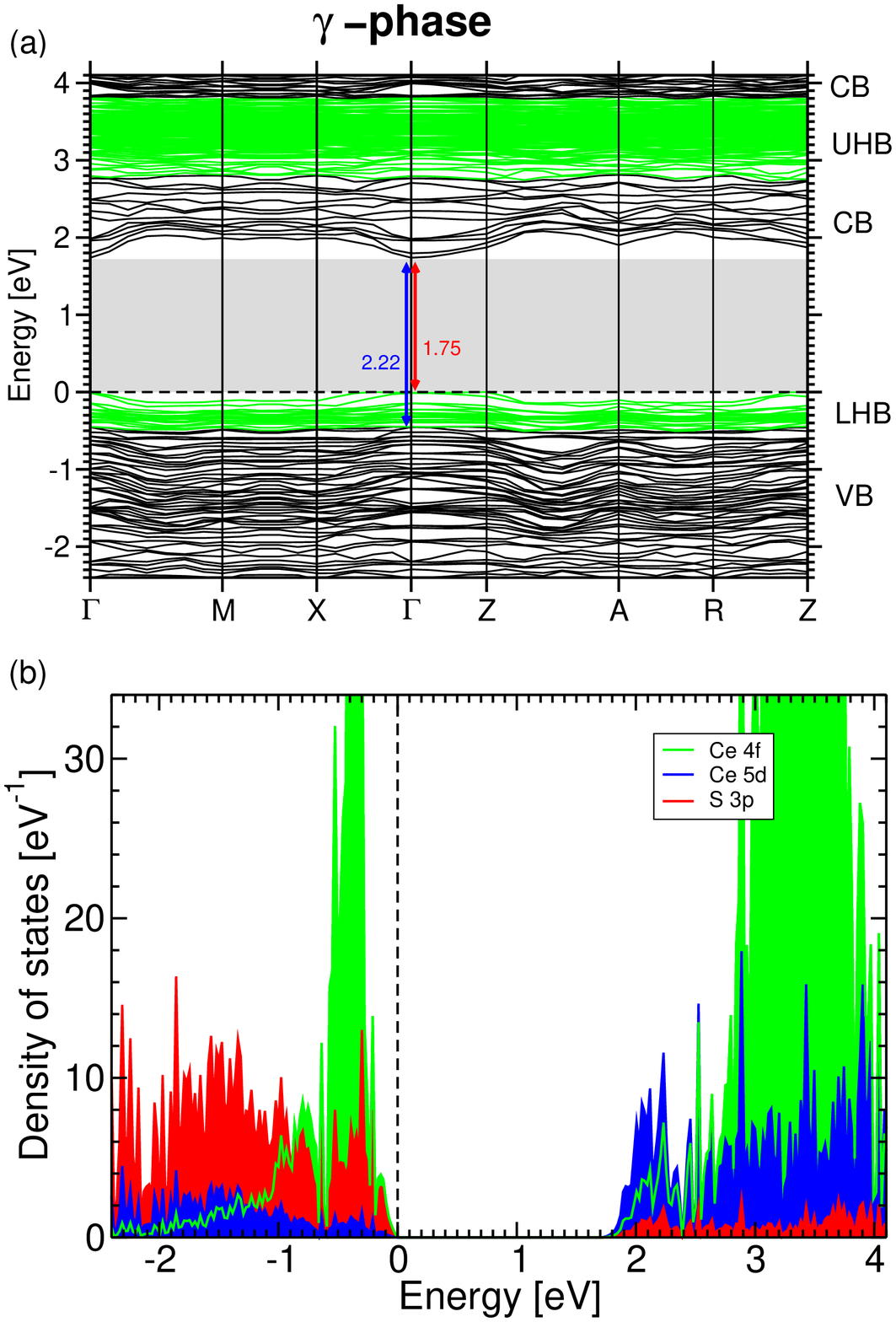}

\caption{\label{fig:gCe32S48fopt-BndsPDOS} (color online) Electronic structure
of pure $\gamma$-Ce$_{2}$S$_{3}$ around the Fermi energy. (a) Band
structure and (b) partial density of states calculated for the fully
optimized system Ce$_{16}$S$_{24}$ (space group $I\bar{4}2d$).
More details are given in the legend of Fig. \ref{fig:aCe2S3fopt.Bands-PDOS}.}
\end{figure}

\begin{flushleft}Figure \ref{fig:gCe32S48fopt-BndsPDOS} depicts the
band structure and PDOS calculated for the fully optimized Ce$_{16}$S$_{24}$
supercell which is the primitive cell of that shown in Fig. \ref{fig:Ce2S3.StrctFeatures}(c).
Comparison with Fig. \ref{fig:aCe2S3fopt.Bands-PDOS} reveals similarities
between the electronic structures of the $\alpha$-phase and the $\gamma$-modification,
but also significant differences which will be addressed in more details
in Sec. \ref{sec:comparison}. Up to an energy of $-$1.0 eV, the vb
is mainly formed by S $3p$-orbitals with small contributions of Ce $4f$-states.
Between $-$1.2 and $-$0.5 eV the vb has increasing admixtures of Ce $4f$-states.
The region of the occupied states above $-$0.5 eV (directly on top of
the vb) is dominated by the LHB with significant components of S $3p$-states.
The difference to the PDOS of the $\alpha$-phase is that the LHB
of $\gamma$-Ce$_{2}$ S$_{3}$ has one maximum only at around $-$0.4
eV. The width of the LHB is around 0.5 eV without any distinct dispersion.
The location of the $4f$ LHB is in good agreement with available
photoelectron spectroscopy data, but inferring its width from these
experiments is difficult given the resolution \citep{witzhugueninlafaitduponttheye794-1996}.\par\end{flushleft}

\begin{flushleft}The single maximum of the LHB in the PDOS is evidence
for cerium cations which are electronically and magnetically equivalent.
This is consistent with the fact that the calculations yield for each
cerium cation occupation numbers of $n_{f}=1.14$. Occupation numbers
larger than one indicate admixtures of other band states, mainly S $3p$-orbitals.
As in the case of $\alpha$-Ce$_{2}$S$_{3}$ the cerium cations of
the $\gamma$-phase have a local effective magnetic moment of 2.6
$\mu_{B}$.\par\end{flushleft}

\begin{flushleft}The cb has predominantly Ce $5d$ character with small
admixtures of S $3p$-states and has its largest dispersion around
$\mathbf{k}=\Gamma$. At the same \textbf{k}-point is the smallest
direct gap between the maximum of the LHB and the cb minimum of around
1.8 eV which is in fact in the range of measured optical gaps between
1.8 and 2.0 eV \citep{witzhugueninlafaitduponttheye794-1996,golubkovprokofievshelykh376-1995,prokofievshelykhgolubkovsmirnov219-1995,dagysbabonas109-1994}.
Thus the present calculations on $\gamma$-Ce$_{2}$S$_{3}$ result
in a dark red color. The rather weak dispersion of both the LHB and
of the Ce $5d$ bands at the bottom of the cb {[}see Fig. \ref{fig:gCe32S48fopt-BndsPDOS}(a)]
implies a rather sharp absorption edge. This is in contrast with the
$\alpha$-allotrope in which both the LHB and the $5d$ cb are more disperse.
The smallest gap between the vb and the cb is 2.3 eV, also at $\mathbf{k}=\Gamma$.
The minimum of the lower edge of the UHB is located approximately
2.8 eV above zero energy.\par\end{flushleft}

\section{\label{sec:comparison}Comparison between the Ce$_{2}$S$_{3}$ modifications}

\begin{figure}
\includegraphics[width=0.9\columnwidth]{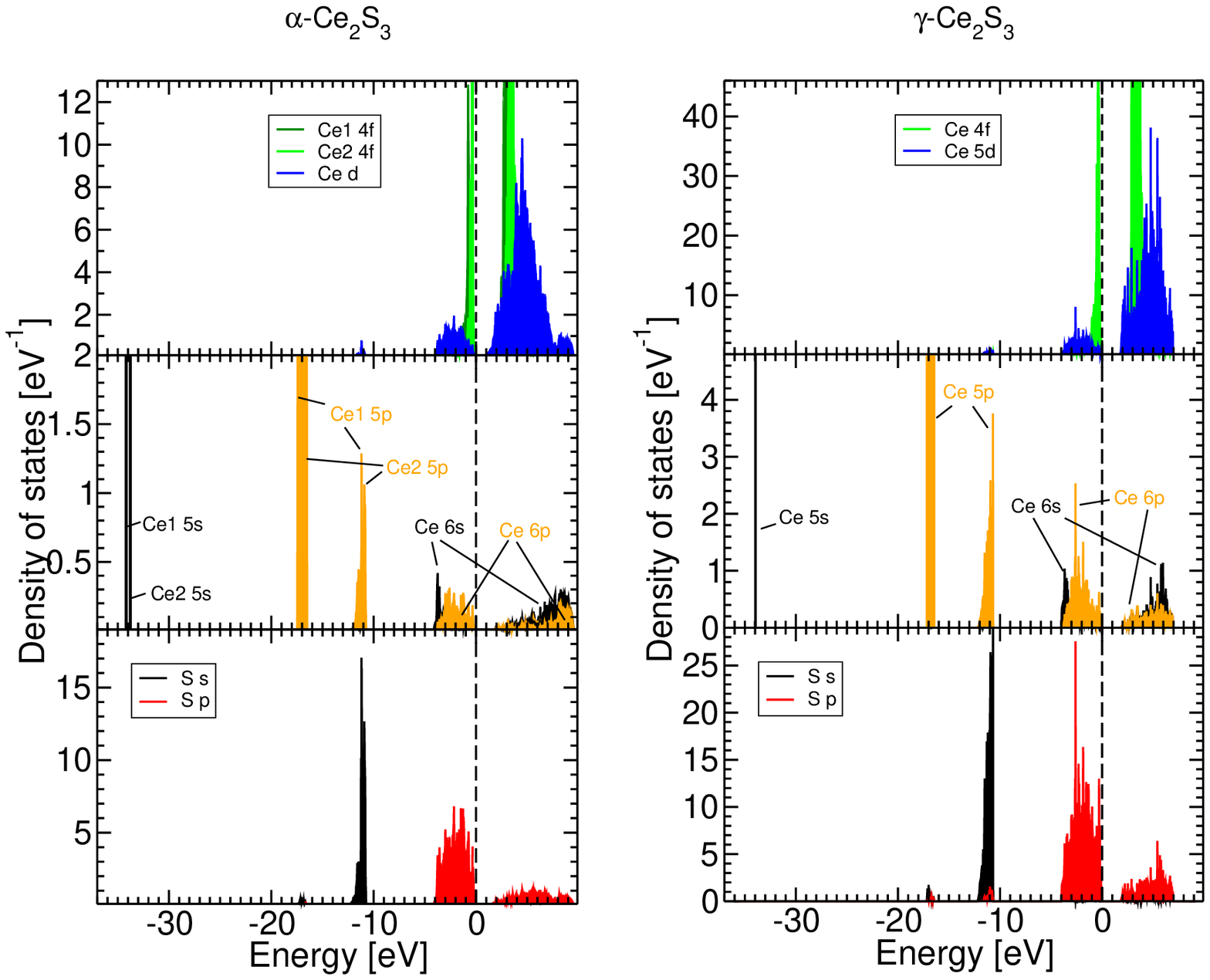}

\caption{\label{fig:allPDOS}(color online) Calculated partial density of
states of two Ce$_{2}$S$_{3}$ modifications in the energy range
between $-$37 and 10 eV.}
\end{figure}

\begin{flushleft}Both allotropes have crystal structures with a relatively
high average coordination of eight sulfur atoms around each Ce atom.
The mass density of the two Ce$_{2}$S$_{3}$ phases calculated from
the optimized crystal structures is similar: 5.10 Mg~m$^{-3}$ ($\alpha$-phase)
and 5.13 Mg~m$^{-3}$ ($\gamma$-phase). For comparison the experimentally
determined mass densities for $\alpha$-Ce$_{2}$S$_{3}$ and $\gamma$-Ce$_{2}$S$_{3}$
are 4.95 and 5.18 Mg~m$^{-3}$ \citep{picondomangeflahautguittardpatrie2-1960}.
The calculations predict that the electronic energy of the $\alpha$-phase
is 6 kJ~mol$^{-1}$ Ce$_{2}$S$_{3}$ lower (i.e. more stable) than
that of the $\gamma$-phase. This is in agreement with the fact that
$\alpha$-Ce$_{2}$S$_{3}$ is the low temperature configuration.
In order to be more stable at higher temperatures the $\gamma$-phase
must have a larger entropy than the $\alpha$-modification. This is
plausible because the $\gamma$-phase contains vacancies on the cationic
sublattice. The voids might cause more low energy lattice phonons
in the $\gamma$-phase than in the $\alpha$-phase. A lowering of
phonon frequencies leads to a higher vibrational entropy. Furthermore,
the distribution of voids in the lattice adds a configurational entropy,
which also tends to stabilize the $\gamma$-phase at elevated temperatures.\par\end{flushleft}

\begin{flushleft}Figure \ref{fig:allPDOS} compares the partial density
of states of $\alpha$-Ce$_{2}$S$_{3}$ and $\gamma$-Ce$_{2}$S$_{3}$
in the energy range of $-$37 to 10 eV. Overall, the electronic structures
of the two allotropes have common features. The bonding between Ce
and S atoms has significant covalent components. It is dominated by
the interaction between Ce $5d$- and S $3p$-orbitals. The centers
of the bonding and anti-bonding $p$-$d$ bands are approximately
7 eV apart in both compounds (\emph{cf.} Fig. \ref{fig:allPDOS}).\par\end{flushleft}

\begin{flushleft}In both modifications, the effective one-particle
energy of this LHB is at the top of the S $3p$ vb. This leads to a
hybridization between Ce $4f$- and S $3p$-orbitals. Electronic excitations
in the visible part of the spectrum involve transitions from the occupied
LHB into the cb, which is dominated by Ce $3d$-levels.\par\end{flushleft}

\begin{flushleft}However, the present calculations reveal a major
distinction between the $\alpha-$ and $\gamma-$ allotropes, due
to the structural differences. The presence of two different coordination
environments of the cerium cations in the $\alpha$-phase causes a
splitting near the minimum of the cb and also within the occupied
$4f$-states at the top of the vb (Figs. \ref{fig:aCe2S3fopt.Bands-PDOS}
and \ref{fig:allPDOS}). As a consequence, the optical transitions
at 1.8 eV responsible for the red color are superimposed by transitions,
which start around 0.8 eV. Furthermore, this distinction between the
two types of Ce cations in the $\alpha$-phase is noticeable also
in the Ce $5s$- and Ce $5p$-semi-core states at $-$34 and $-$17 eV, respectively
(Fig. \ref{fig:allPDOS}). Both states are split by about 0.5 eV whereby
the $5s$ and $5p$-states of Ce1 are located at lower energies than
that of Ce2. This implies that the effective potential for electrons
around Ce1 is more attractive than around Ce2.\par\end{flushleft}

\section{Summary and conclusions}

\begin{flushleft}The present GGA+U calculations yield a very good
description of the crystal structures and of the chromatic properties
of $\alpha$-Ce$_{2}$S$_{3}$ and $\gamma$-Ce$_{2}$S$_{3}$ involving
$4f$-electrons explicitely. With a Coulomb repulsion parameter of
$U'=$4 eV the calculated cell volume and most of the interionic distances
for both Ce$_{2}$S$_{3}$ phases are less than 1\% larger than that
determined from x-ray diffraction measurements. This is comparable
to the accuracy of standard GGA calculations as applied to systems
without $f$-electrons. This is a major improvement over standard
GGA calculations, which underestimate the cell volume of Ce$_{2}$S$_{3}$
derivatives by as much as 6\% and fail to predict their semiconducting
character \citep{windikspourovskiygeorgeswimmer-2006}.\par\end{flushleft}

\begin{flushleft}With the same $U$' parameter the present calculations
yield effective one-electron energies, which allow a quantitative
interpretation of optical transitions and thus accounts for the colors
of the two Ce$_{2}$S$_{3}$ allotropes in a satisfactory manner.
The vb and cb of both allotropes are dominated by S $3p$ and Ce $5d$-states,
respectively. In both phases the occupied Ce $4f$-states are energetically
located between the vb and cb, whereby the smallest gap between the
occupied Ce $4f$- and the empty $5d$-states ($\Delta_{4f-5d}$) is
direct and located at $\mathbf{k}=\Gamma$.\par\end{flushleft}

\begin{flushleft}For the $\alpha$-phase the calculations yield $\Delta_{4f-5d}=$
0.8 eV and a rather small density of states in the range of 0.8--1.5
eV above the vb. The small density of states is due to three low-lying
unoccupied bands with strong dispersion and corresponds to low absorption
of radiation. The dispersion is related to the (Ce1)S$_{8}$ coordination
polyhedron and it is the reason for the smaller optical band gap and
explains the brownish-black hue $\alpha$-Ce$_{2}$S$_{3}$. The dark
red color (burgundy) of the $\gamma$-Ce$_{2}$S$_{3}$ is due to
$\Delta_{4f-5d}$=1.8 eV and a rather steep increase of the density
of states around the lower edge of the cb. The $\gamma$-allotrope
is charcaterized by somewhat less dispersive bands and hence a sharper
absorption edge.\par\end{flushleft}

\begin{flushleft}Thus, the present calculations provide an explanation
for the difference of 1 eV in the optical gaps of $\alpha$-Ce$_{2}$S$_{3}$
and $\gamma$-Ce$_{2}$S$_{3}$. The main factor is the presence of
two types of Ce cations in the $\alpha$-phase with two different
coordination environments and the related difference of 3\% between
the average Ce--S bond distances in the two allotropes. Despite the
different sulfur environments the calculations yield for cerium cations
effective local magnetic moments of 2.6 $\mu_{B}$ which is in agreement
with magnetic measuremts.\par\end{flushleft}

\begin{flushleft}It is gratifying that the computed electronic energy
of the low-temperature modification ($\alpha$-Ce$_{2}$S$_{3}$)
is lower by 6 kJ~mol$^{-1}$ (per formula unit) than that of the
high-temperature modification ($\gamma$-Ce$_{2}$S$_{3}$). At low
temperatures, the Gibbs free energy, which determines phase stabilities,
is therefore lower for the $\alpha$-phase. At elevated temperatures,
the presence of vacancies in the $\gamma$--phase is expected to enhance
the vibrational and configurational entropy, thus stabilizing the
high-temperature phase.\par\end{flushleft}

\begin{flushleft}The present GGA+U approach offers a satisfactory
description of the Ce $4f$-states in terms of structural and optical
properties. However, the introduction of the on-site Coulomb repulsion
does not cure the drawback of standard DFT methods of underestimating
the gap between the itinerant S $3p$- and the Ce $5d$-band states.
The smallest calculated direct gaps between the vb and the cb of $\gamma$-Ce$_{2}$S$_{3}$
is computed to be 2.3 eV. This is approximately 1 eV smaller than
the value determined from optical, photoconductivity, and photoluminescence
spectra. The same band gap calculated for $\alpha$-Ce$_{2}$S$_{3}$
is about 1.3 eV and probably also underestimated by 1 eV. A remedy
of this problem might be the combination of GGA+U and GW techniques
\citep{shishkinkresse74-2006}.\par\end{flushleft}

\begin{flushleft}In conclusion, the present work demonstrates that
the GGA+U approach is successful in describing quantitatively the
structural and electronic properties of $\alpha$-Ce$_{2}$S$_{3}$
and $\gamma$-Ce$_{2}$S$_{3}$. The GGA+U approach overcomes a major
shortcoming of standard GGA band structure calculations, which incorrectly
yields a metallic bandstructure with $4f$-states at the Fermi level.
Instead, we found Ce$_{2}$S$_{3}$ to be an $f$-electron Mott insulator,
in which strong correlation effects open up a Mott gap within the
$4f$-states, between a LHB and an UHB. The LHB is found to play a
key role in the physical properties of these compunds, since the optical
transitions responsible for their color are between the top of the
LHB and the bottom of the $5d$ cb. Furthermore, subtle differences
in the calculated electronic structure of the two allotropes (most
notably the presence of two inequivalent cerium sites in the $\alpha$-modification,
as well as more dispersive bands) give insight into the differences
in the optical properties and color of these materials. The present
GGA+U methodology applied here to RE compounds, could also be used
to predict structural and opto-electronic properties of materials
where RE elements are used as dopants. This capability will be particularly
useful in the design and improvements of materials for lighting and
display technologies with major benefits for our society and the environment.\par\end{flushleft}

\begin{ack}
This work has been supported by the European Commission within its
programme Research and Training Network {}``Psi-k f-electrons''
Contract HPRN-CT-2002-00295, as well as by Ecole Polytechnique and
CNRS. The authors are grateful to the other participants of this Network
and to Axel Svane for coordination. The authors are grateful to Paul
Saxe, Walter Wolf, Alexander Mavromaras, and Benoit Leblanc (Materials
Design) and to Jan Tomczak, Alexander Poteryaev (Ecole Polytechnique)
for useful discussions. 
\end{ack}

\end{document}